\documentstyle[11pt,newpasp,twoside,epsf]{article}
\def\edcomment#1{\iffalse\marginpar{\raggedright\sl#1\/}\else\relax\fi}
\reversemarginpar

\begin{document}
\title{Evolution of binaries producing Type Ia supernovae,
luminous supersoft X-ray sources, and recurrent novae
} 

\author{Izumi Hachisu} 

\affil{College of Arts and Sciences, University of Tokyo, 
Meguro-ku, Tokyo 153-8902, Japan}

\begin{abstract}
We have been proposing two evolutionary paths to Type Ia supernovae
(SNe Ia), which are called the supersoft X-ray source (SSS) channel 
and the symbiotic channel, depending on the orbital separation 
just prior to an SN Ia explosion.
The essential difference of our treatment is inclusion of strong,
optically thick winds on mass-accreting white dwarfs (WDs) 
in the elementary processes of binary evolution when the mass 
accretion rate on to WDs exceeds a critical rate of 
$\dot M_{\rm cr} \sim 1 \times 10^{-6} M_\odot$ yr$^{-1}$.
Once optically thick winds begin to blow from the WD,
the binary can avoid forming a common envelope in some cases.  
We call this {\it accretion wind}.  So that the WDs are able 
to grow up to the Chandrasekhar mass and explode as an SN Ia, 
showing SSS or recurrent nova phenomena in the way to SNe Ia. 
Thus, the accretion wind process of WDs can open new channels 
to SNe Ia.  We have modeled the LMC supersoft source RX J0513.9-6951 
as an example of the systems in the accretion wind phase.
Further inclusions of two other elementary 
processes make the channels much wider; these are
the case BB mass transfer in the SSS channel 
and the strong orbital shrinkage during the superwind phase of the 
primary star in the symbiotic channel.  As a result, the estimated 
birth rate of SNe Ia via these two channels becomes compatible with 
the observation in our Galaxy.  Interestingly, the U Sco and T CrB 
subclasses of recurrent novae can be naturally understood as a part
of evolutionary stages in these two SSS and symbiotic channels 
to SNe Ia, respectively.   Thus we have a unified picture
of binary evolutions to SNe Ia, luminous SSS, and recurrent novae. 
\end{abstract}

\section{Introduction}
     Type Ia supernovae (SNe Ia) are one of the most luminous 
explosive events of stars.  Recently, SNe Ia have been used 
as good distance indicators that provide a promising tool 
for determining cosmological parameters (Perlmutter et al. 1999;
Riess et al. 1998) 
because of their almost uniform maximum luminosities.
However, one of the most important unresolved problems on SNe Ia is
that we do not know the exact progenitor systems of SNe Ia.
\par
    It is widely accepted that the exploding star itself is 
an accreting white dwarf (WD) in a binary
(Nomoto 1982; Nomoto,  Thielemann, \& Yokoi 1984).
However, the companion star (and thus the observed binary system)
is not known.  Several objects have ever been considered, 
which include merging double white dwarfs
(Iben \& Tutukov 1984; Webbink 1984),
recurrent novae (Starrfield, Sparks, \& Truran, 1985),
symbiotic stars (Munari \& Renzini 1992) etc 
(Livio 2000, for recent summary). 
\par
     Among these models, the systems of double white dwarfs 
(the double degenerate model) are not theoretically
supported; when two WDs merge, carbon ignites not at the center
but near the surface layer of the C+O WD; 
so that the C+O WD is converted into an O+Ne+Mg WD (Saio \& Nomoto 1998).  
If the total mass of the merger exceeds the Chandrasekhar mass,
the merged WD does not explode as an SN Ia but collapses to a neutron star 
(Nomoto \& Kondo 1991).  

\begin{figure}
\plottwo{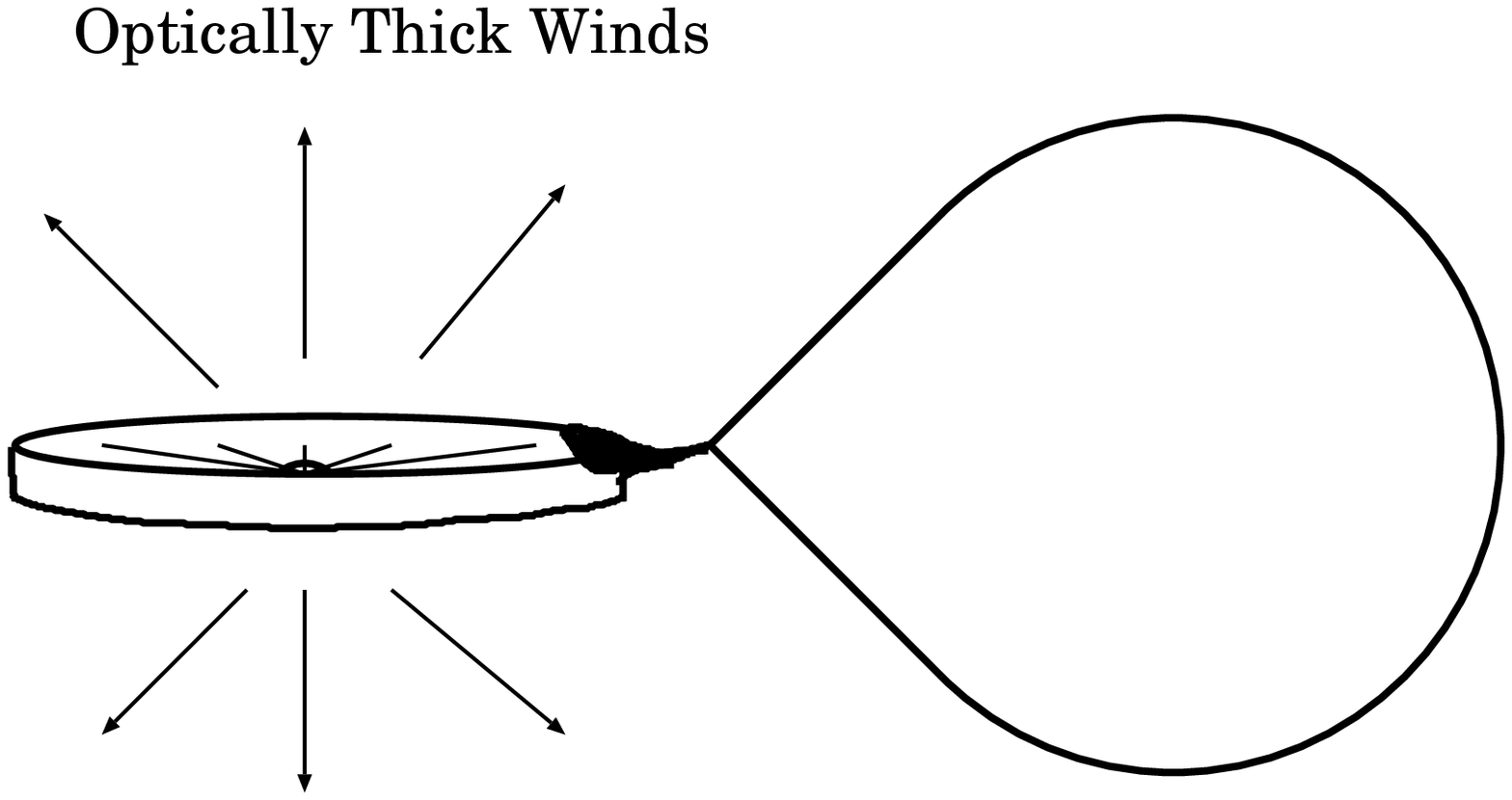}{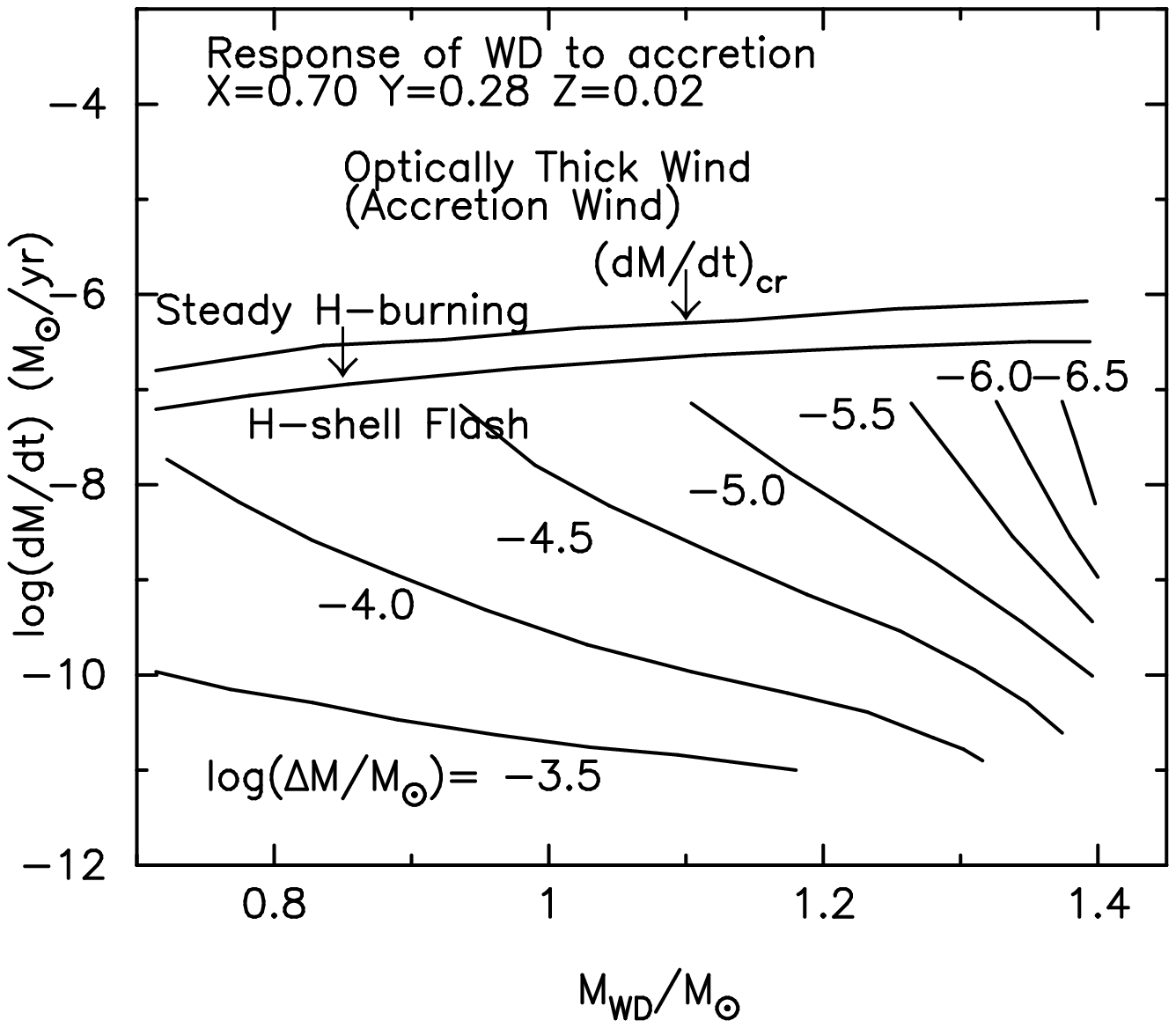}
\caption{{\bf Left:} 
Optically thick winds blow from mass-accreting white dwarfs
when the mass transfer rate exceeds a critical 
rate, i.e., $\dot M_{\rm acc} > \dot M_{\rm cr}$.  Here, we assume 
that the white dwarf accretes mass from equatorial region and, 
at the same time, blows winds from polar regions.  {\bf Right:} 
Response of white dwarfs to mass accretion is illustrated
in the white dwarf mass and the mass accretion rate plane,
i.e., in the $M_{\rm WD}$-$\dot M_{\rm acc}$ plane.
Strong optically thick winds blow above the line of
$\dot M_{\rm acc} > \dot M_{\rm cr}$.  
There is no steady state burning below 
$\dot M_{\rm acc} < \dot M_{\rm std}$.  Instead, intermittent
shell flashes occur.  The envelope mass at which hydrogen ignites
to flash is also shown (taken from Fig. 9 of Nomoto 1982).
}
\end{figure}

     On the other hand, the single degenerate systems has become more
promising, in which a C+O WD gains mass from a non-degenerate 
companion star and grows its mass up to the Chandrasekhar mass.
Such promising progenitors include luminous supersoft X-ray sources
(SSS), recurrent novae (RNe), and symbiotic stars.   
However, the most important problem in these single degenerate systems 
is that we did not know the mechanism how to grow the mass of the WDs.
Instead, there are effective mass-losing mechanisms 
such as nova ejection 
and common envelope evolutions that get rid of the mass accumulated 
on the WD.  Here, I review and summarize our recent developments
of the mechanisms that enable to grow the mass of WDs in the 
single degenerate systems and give a unified picture of
evolutions of SSSs, recurrent novae, and SNe Ia.

\section{Accretion wind evolution}
     First, we briefly explain how the mass of WDs in the single
degenerate systems grow toward the Chandrasekhar mass limit.
The reason that the standard binary evolution theory 
fails to explain the evolutionary path to SNe Ia and recurrent novae 
lies mainly in the theoretical prediction of a common envelope
formation and the ensuing spiral-in process, because these processes 
inhibit the growth of white dwarfs in binary systems.  As discussed
in many previous papers on binary evolution, 
it has been widely accepted that a hydrogen-rich envelope on 
a mass-accreting white dwarf expands to a red giant size 
when its mass accretion rate, 
$\dot M_{\rm acc}$, exceeds a critical limit, i.e.,
\begin{equation}
\dot M_{\rm cr} = 8.3 \times 10^{-7} 
({{M_{\rm WD}} \over {M_\odot}} - 0.4) ~M_\odot 
\mbox{~yr}^{-1},
\end{equation}
(see, e.g., Fig. 9 of Nomoto 1982)
and easily forms a common envelope.
Once a common envelope is formed, two stars begin to spiral-in 
each other as a result of viscous drag, thus producing
a double degenerate system (Iben \& Tutukov 1984; Webbink 1984).
\par
     However, Hachisu, Kato, \& Nomoto (1996)
found that white dwarfs begin to blow 
strong winds ($v_{\rm wind} \sim 1000$ km s$^{-1}$ and
$\dot M_{\rm wind} > 1 \times 10^{-6} M_\odot$ yr$^{-1}$)
when $\dot M_{\rm acc} > \dot M_{\rm cr}$,
as illustrated in Fig. 1, thus 
preventing the binary from collapsing.
The binary does not spiral-in but keeps its separation. 
Hydrogen burns steadily, and therefore the helium layer of 
the white dwarf can grow at a rate of 
$\dot M_{\rm He} \approx \dot M_{\rm cr}$.
The other transferred matter is blown in the wind 
($\dot M_{\rm wind} \approx \dot M_{\rm acc} - \dot M_{\rm cr}$),
where $\dot M_{\rm wind}$ is the wind mass loss rate. 
We call this {\it accretion wind} because it begins to blow 
when the accretion rate exceeds the critical rate.
\par
     Therefore, we have to revise Fig. 9 of Nomoto (1982),
which has been used widely in binary evolution scenarios to show the 
basic processes of mass-accreting white dwarfs.
The revision to Nomoto's Fig. 9 is shown in Fig. 1.
The most important difference of Fig. 1 from 
Nomoto's Fig. 9 is that white dwarfs increase their masses 
in a much wider parameter region. 
In our new diagram, the status of a mass-accreting white dwarf 
is specified by the following three phases:
(1) accretion wind phase ($\dot M_{\rm acc} > \dot M_{\rm cr}$); 
(2) steady shell-burning phase 
($\dot M_{\rm std} < \dot M_{\rm acc} < \dot M_{\rm cr}$);
and (3) intermittent shell-flash phase
($\dot M_{\rm acc} < \dot M_{\rm std}$), where
$\dot M_{\rm std}$ means the lowest limit of the mass accretion rate
for steady hydrogen shell-burning.  
The new growing region of white dwarfs,
$\dot M_{\rm acc} > \dot M_{\rm std}$, 
is much wider than old Nomoto's (1982) narrow strip,
$\dot M_{\rm std} < \dot M_{\rm acc} < \dot M_{\rm cr}$.

\begin{figure}
\plottwo{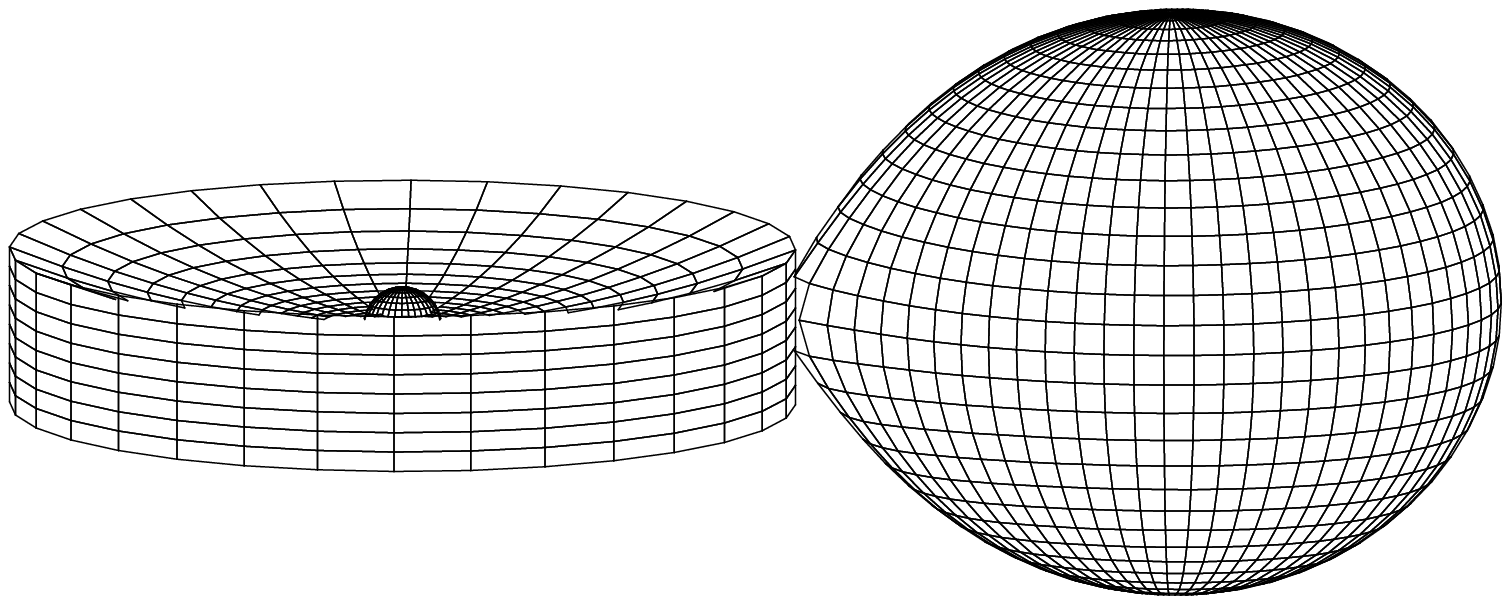}{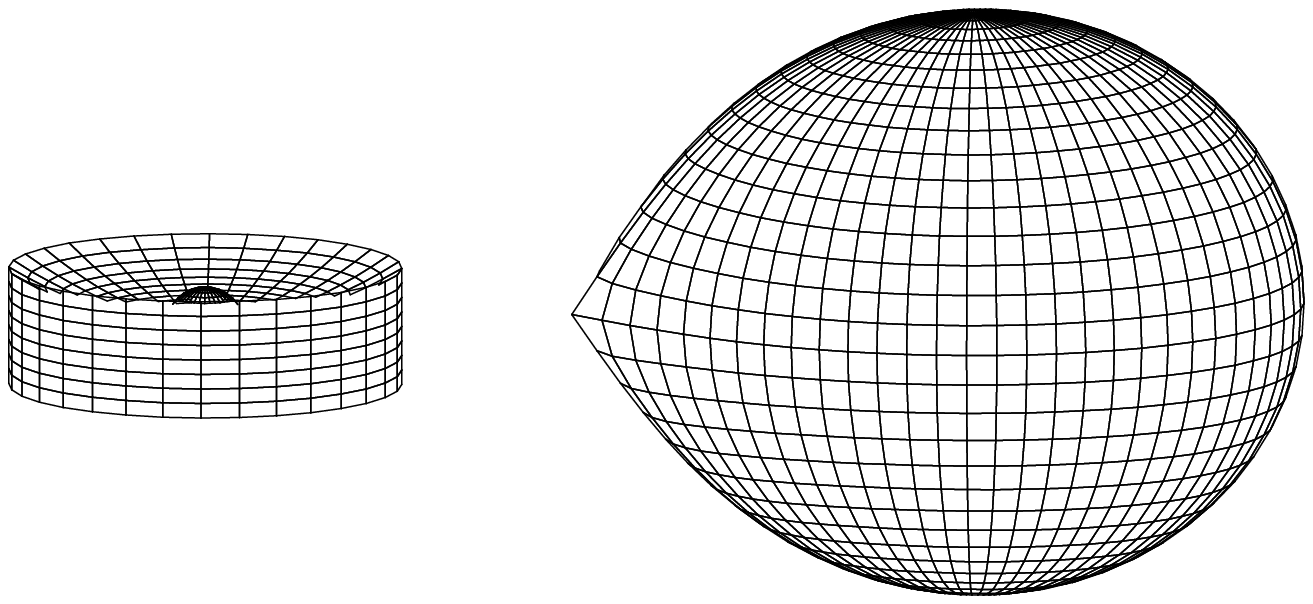}
\caption{{\bf Left:}
Geometrical configuration of our RX J0513.9-6951 model in the optical 
high state (wind phase).  The cool component is a slightly evolved 
MS companion filling up its inner critical Roche lobe.
The hot component, WD, blows a strong wind.
The accretion disk is blown off outward and has a larger size than
the no-wind phase.  {\bf Right:}
Geometrical configuration in the optical low state (no-wind phase).
}
\end{figure}

\begin{figure}
\plotone{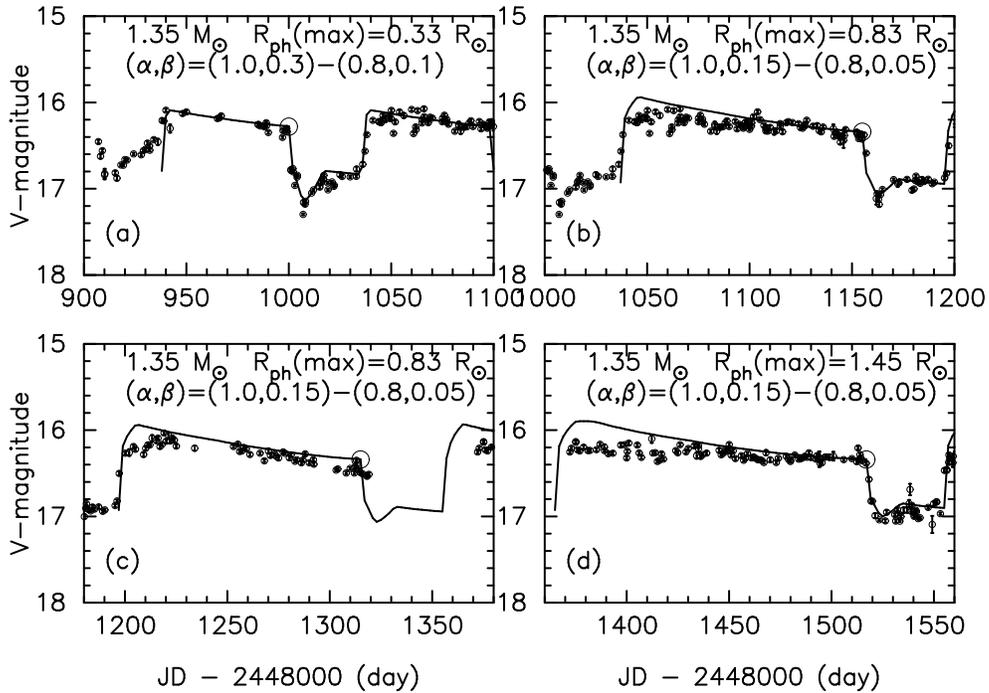}
\caption{
Light curve is plotted against time for a $1.35 M_\odot$ white 
dwarf model together with the MACHO data (Alcock et al. 1996). 
We assume a larger size of the accretion disk 
(1.0 times the Roche lobe size) in the optical high state
(the wind phase) but a smaller size (0.8 times the Roche lobe size)
in the optical low state (the static, no-wind phase). 
The transition from the optical high to low state 
corresponds to the cease point of the strong wind 
(large open circles).  
}
\end{figure}

\subsection{RX J0513.9-6951 as an example of accretion wind phase}
As an example of the {\it accretion wind} phase, 
we have modeled long-term optical light curves of the LMC supersoft 
X-ray source RX J0513.9-6951 based on our wind model of 
mass-accreting white dwarfs.  The source shows a quasi-periodic 
variability, the period of which is $\sim 170$ d and 
the amplitude of which is $\sim 1$ mag (Alcock et al. 1996). 
The source has a relatively longer ($\sim 130$ d) optical high-state 
($V \sim 16.2$) and a shorter ($\sim 40$ d) optical low-state 
($V \sim 17.0$) with the transition time of several days.  
X-rays were observed only in the optical low-states.  
This indicates that a strong wind blows only in the optical high-state 
because X-rays are self-absorbed by the wind.  
To examine the idea of the strong wind, 
we have calculated envelope solutions of mass-accreting WDs 
and found that optically thick winds really occur when the mass 
accretion rate exceeds a critical rate of  $\dot M_{\rm cr}$
for $M_{\rm WD} > 0.7 M_\odot$ 
even in a low metallicity of $Z=0.004$ of LMC. 
\par
     We attribute the $100-150$ d bright 
phase to our strong wind phase with an expanded white dwarf 
photosphere, whereas the $\sim 40$ d 
less bright phase to our static (no-wind) phase with a shrinking 
envelope of the WD.  The duration and 
the brightness of the optical high state is explained by 
the model of a WD mass, $M_{\rm WD}= 1.35 ~M_\odot$, 
with an extending accretion disk edge over the Roche lobe size
during the wind phase as illustrated in Fig. 2. 
Irradiation of the accretion disk by the WD photosphere
plays an essential role to reproduce the optical light curve 
because a large part of the optical light comes from the 
accretion disk. 
\par
     The switching mechanism from the optical high to low state 
is understood as follows: 
Strong winds from the WD themselves make the separation increase, 
because the high velocity wind do not carry away large angular momentum.
The winds further strip off the very surface layer of the companion star.
As a result, the mass transfer rate decreases and it finally 
stops the strong winds.  After the wind stops, the surface of
the companion recovers in a thermal timescale of the envelope
and the mass transfer rate increases up and strong winds begin to blow
again.  Thus, the semi-periodic cycles repeat.
\par
     Light curve fittings indicate that the optical high state 
is initiated by a sudden increase in the mass transfer rate of
$\dot M_{\rm acc} \sim 1 \times 10^{-5} ~M_\odot$. 
Then, the envelope of the white dwarf expands to 
$R_{\rm ph} \sim 0.1 ~R_\odot$, not exceeding over the 
accretion disk size of $\sim 2 ~R_\odot$.  Our light curve
fittings are shown in Fig. 3.  In our models,
about 40\% of the transferred matter can be accumulated on
the white dwarf but the rest is blown off in the wind. 
This suggests that it takes $\sim 1 \times 10^5$ yr for the white dwarf 
of $1.35 ~M_\odot$ to grow to $1.38 ~M_\odot$ 
and to explode as a Type Ia supernova.

\subsection{Two evolutionary paths to RNe and SNe Ia}
     Based on the new mechanism of the accretion wind evolution as 
shown in Fig. 1, Hachisu, Kato, \& Nomoto (1999a) and Hachisu et al. (1999b) 
have proposed two paths to SNe Ia (see also, Li \& van den Heuvel 1997).
In these two paths, WDs can accrete mass continuously from 
the companion and grow at a rate of $\dot M_{\rm cr}$ of eq. (1).
Therefore, candidates for SN Ia progenitors are systems at the final
stages on these paths; i.e., one is a supersoft
X-ray source (SSS) consisting of a WD and a lobe-filling,
more massive, main-sequence (MS) companion as shown in the late stage of
Fig. 4 (the WD+MS systems; see also Li \& van den Heuvel 1997)
and the other is a symbiotic star consisting of a WD
and a mass-losing red giant (RG) as shown in the late stage of Fig. 5 
(the WD+RG systems, see also Hachisu et al. 1996). 
Both the systems contribute to main parts of the SN Ia birth rate
(Hachisu et al. 1999a, 1999b).
\par
     Hachisu et al. (1999a, 1999b) have followed many binary evolutions
until SN Ia explosions.  The final outcome is summarized 
in Fig. 6 for the initial orbital period, $P_0$ (in units of days),
and the initial donor mass, $M_{d,0}$, which is representing 
$M_{\rm RG,0}$ (the initial mass of the red-giant component), 
or $M_{\rm MS,0}$ (the initial mass of the slightly 
evolved main-sequence component).  In this figure,
the initial mass of the white dwarf is assumed to be 
$M_{\rm WD,0}= 1.0 M_\odot$.  
The right-hand region, longer orbital periods of
$P_0 \sim 100$---1000 d, represents 
our results for the WD+RG system. 
In the left-hand region, shorter orbital periods of $P_0 \sim 0.3$---10 d, 
are shown the results for the WD+MS system.
\par
     We plot the initial parameter regions of binaries (thin solid 
curves in Fig. 6) that finally produce an SN Ia, which are corresponding to 
stage (F) in Fig. 4 or stage (D) in Fig.5.
Left diagram in Fig. 6 shows such two regions of the WD+MS systems 
(SSS channel) and the WD+RG systems (symbiotic channel).  We have added 
in this figure the final parameter regions (thick solid curves)
just before an SN Ia explosion occurs.  Starting from the initial region
encircled by the thin solid curves, binary systems evolve and 
can explode as an SN Ia in the regions enclosed 
by the thick solid curves.  Hachisu \& Kato (2001)
showed that some recurrent novae lie on the boarder of their 
parameter region that can produce an SN Ia 
(see also Kato 2001 in this volume).
Among six recurrent novae with known orbital periods,
five fall in the final regions of SN Ia progenitors 
just before the explosion.  Only T Pyx is far out of the regions 
of SNe Ia.

\begin{figure}
\plottwo{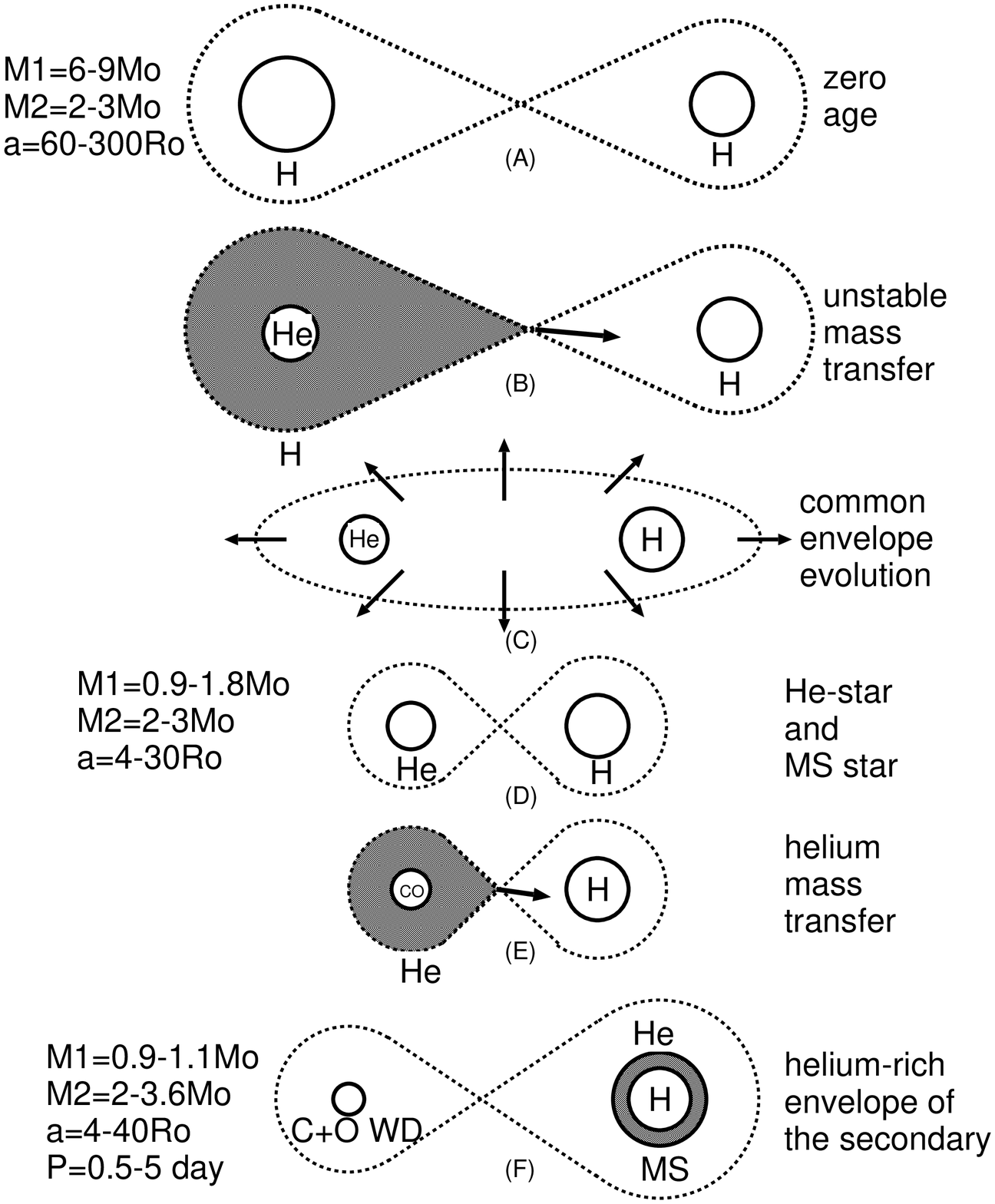}{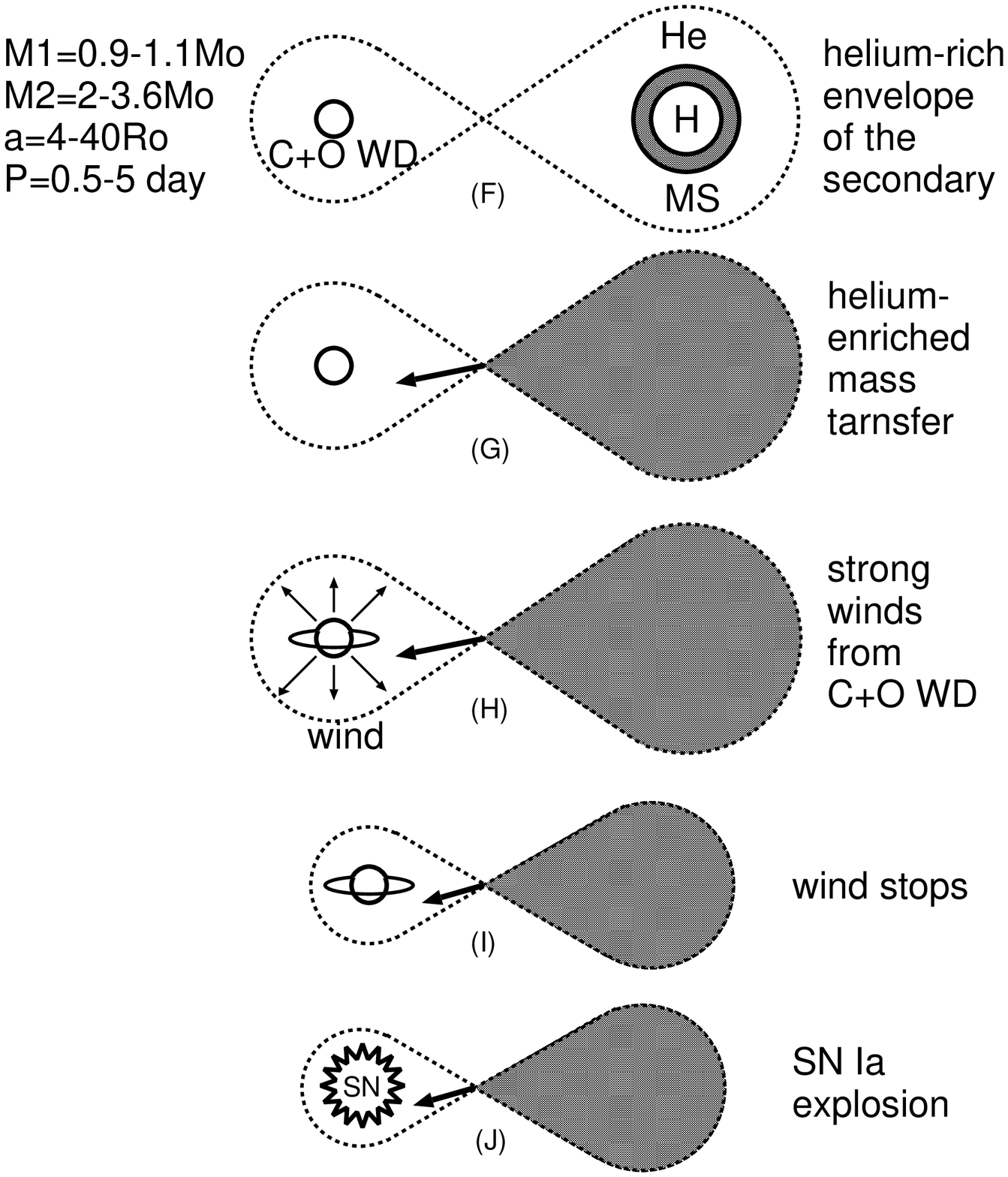}
\caption{ SSS channel.  {\bf Left:} 
Early evolutionary path through the common envelope evolution to
the helium (Case BB) mass transfer.  {\bf Right:}  
Late evolutionary path to an SN Ia explosion in our wind model.
}
\end{figure}

\begin{figure}
\plottwo{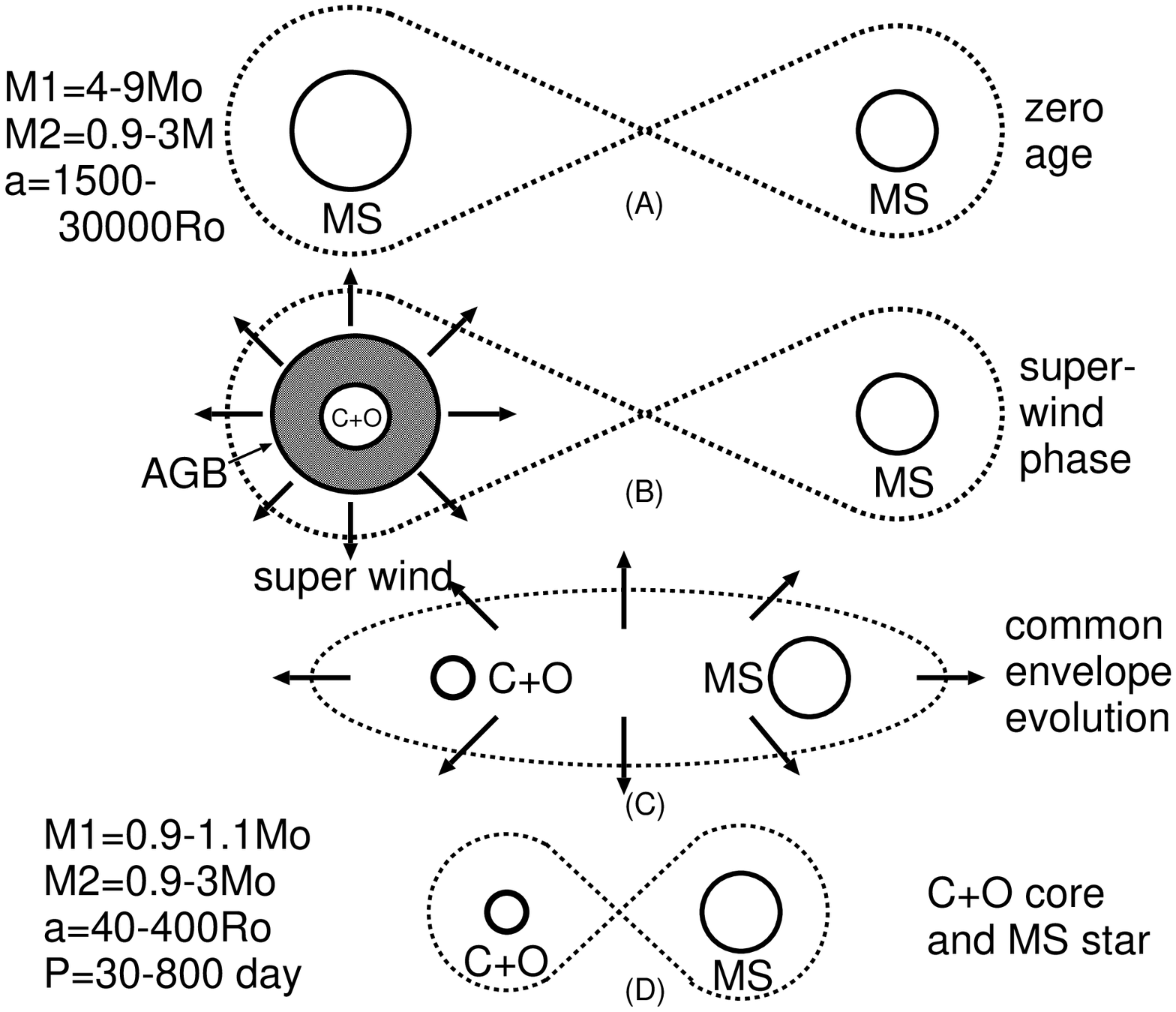}{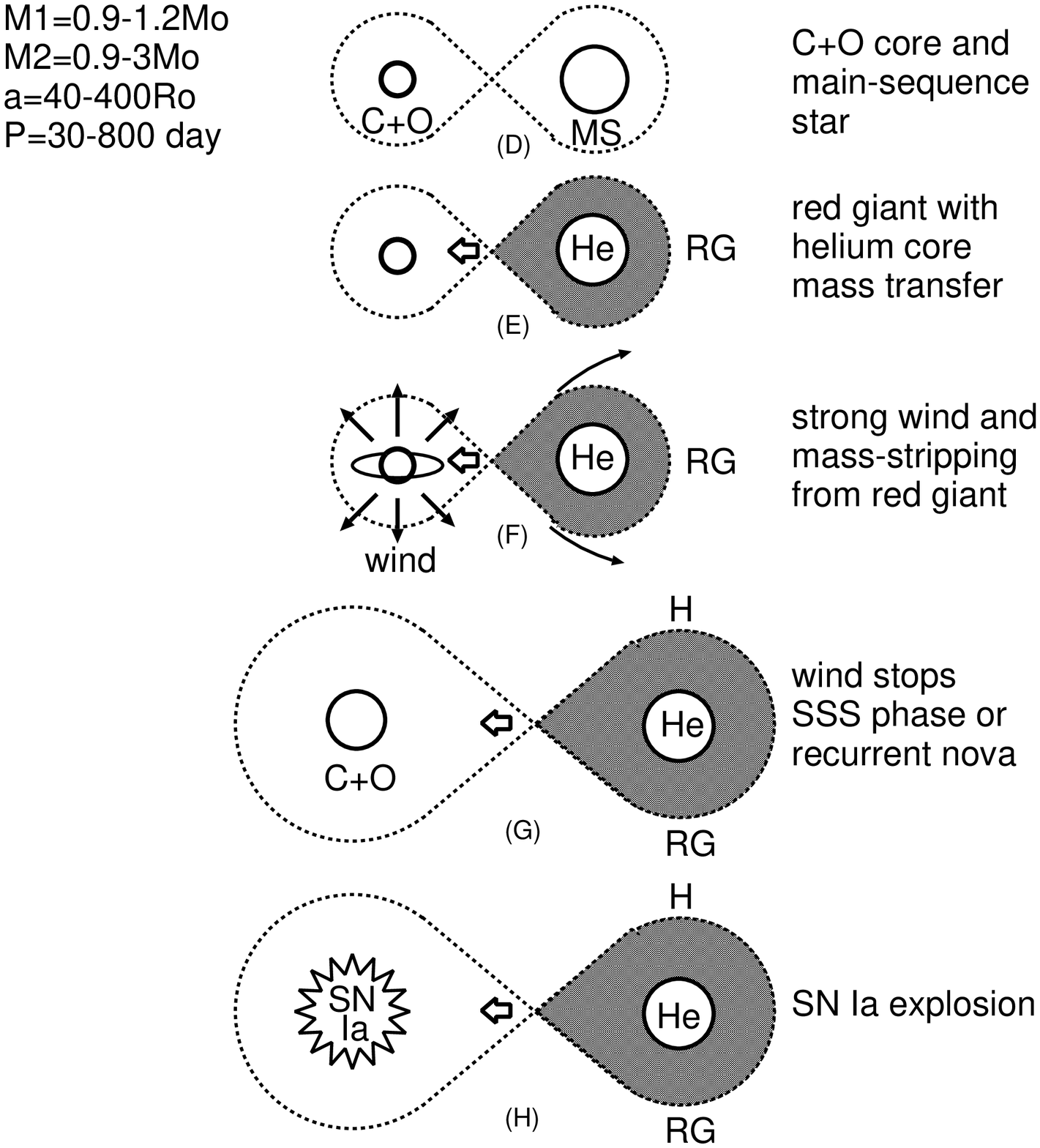}
\caption{ Symbiotic channel.
{\bf Left:} Early evolutionary path through the first common
envelope evolution to a symbiotic system. 
{\bf Right:} Late evolutionary path through 
the symbiotic channel to an SN Ia.
}
\end{figure}

\section{Case BB mass transfer}
      It has been shown that the ejecta of the recurrent nova
U Sco are helium-rich as much as He/H$\sim 2$ by number
(Williams et al. 1981).  The accretion disk of U Sco is also
suggested to be helium-rich (Hanes 1985).  Many of the SSSs 
have a stronger He {\scriptsize II} line than Balmer lines, 
thus also suggesting that the companion of the WD+MS systems 
(SSS channel) is helium-rich.  
In this section, therefore, we describe 
the reason that the cool component of the WD+MS systems has 
a helium-rich envelope.
\par
     Previous works consider only the initial systems 
consisting of a more massive AGB star with a C+O core and a less massive 
main-sequence star.  This system undergoes a common envelope evolution 
and it finally yields a binary system of a mass-accreting C+O WD and 
a lobe-filling MS/subgiant star.   Thus, they do not
include another important evolutionary path, in which
a more massive component fills up its inner 
critical Roche lobe when it develops a {\it helium core} 
of $\sim 0.8-2.0 M_\odot$ in its {\it red giant phase} 
as shown in the early stage of Fig. 4.
\par
     We consider, for example, a close binary with the primary mass of 
$M_{1,i} = 7 M_\odot$ and the secondary mass of $M_{2,i} = 2 M_\odot$
with the initial separation of $a_{i} \sim 80-600 ~R_\odot$.
When the primary has evolved to a red giant developing a helium 
core of $M_{\rm 1,He} \sim 1.0-1.6 M_\odot$, 
it fills up its inner critical Roche lobe.
Then, the binary undergoes a common envelope (CE) evolution, 
which yields a much more compact close binary 
of a naked helium star of $M_{1,{\rm He}} \sim 1.0-1.6 M_\odot$ and 
a main-sequence star of $M_2= 2 M_\odot$ 
with the final separation of $a_{\rm f, CE} \sim 4-50 ~R_\odot$. 
Here, we assume a relation of  
\begin{equation}
{{a_{\rm f,CE}} \over {a_i}} \sim \alpha_{\rm CE} 
\left({{M_{\rm 1,core}} \over {M_{1,i}}} \right) 
\left({{M_2} \over {M_{1,i}-M_{\rm 1,core}}} \right),
\end{equation}
with the efficiency factor of $\alpha_{\rm CE}=1.0$ for the
common envelope evolutions.
\par
     After the common envelope evolution, 
the radii of the inner critical Roche lobes of the primary and 
the secondary become $R_1^* \sim 0.36 a_{\rm f,CE}$ and 
$R_2^* \sim 0.4 a_{\rm f,CE}$, respectively.
Since the secondary radius should be smaller than its inner critical 
Roche lobe, i.e., $1.7 R_\odot \le R_2 < R_2^*$
($1.7 R_\odot$ is the radius of a $2 M_\odot$ ZAMS star), 
the initial separation $a_{i}$ should exceeds $80 ~R_\odot$.  
The upper bound of the initial separation 
is obtained from the maximum radius of the $7 M_\odot$ star 
which has formed a helium core, i.e., 
$a_{i} \approx 2 R_{\rm 1,max} <~ 2 \times 300 ~R_\odot$. 
Thus, the allowable range of the initial separations is
$80 ~R_\odot <~ a_{i} <~ 600 ~R_\odot$ for a pair
of $M_{1,i}= 7 M_\odot$ and $M_{2,i}= 2 M_\odot$. 
Then, we have $R_1^* \sim 1.5-18 ~R_\odot$ and $R_2^* \sim 1.7-20 ~R_\odot$
after the first common envelope evolution.
\par 
     After the hydrogen-rich envelope is stripped away and 
hydrogen shell burning vanishes, the naked helium core 
contracts to ignite a central helium burning and 
becomes a helium main sequence star.  
When the helium star forms a C+O core, its helium envelope 
expands to fills its inner critical Roche lobe again.  
The helium is transferred stably to the secondary 
on an {\it evolutionary time scale} of $\tau_{\rm EV} \sim 10^5$ yr
because the mass ratio is smaller than $0.79$ ($q= M_1/M_2 < 0.79$).
The resultant mass transfer rate is 
$\dot M_1 \sim 10^{-5} M_\odot$ yr$^{-1}$. 
No common envelope is formed for such a low rate.
After the helium envelope is lost, the primary becomes a C+O WD 
of $M_{\rm C+O} \sim 0.9-1.1 M_\odot$ and the separation increases 
by 10\%$-$40\%, i.e., 
$a_{\rm f, He} \sim (1.1-1.4) a_{\rm f, CE} \sim (4-70) R_\odot$. 
Here, we assume the conservations of the total mass and angular momentum 
during the helium mass transfer
to obtain the separation after the helium mass transfer, 
$a_{\rm f, He}$.
The secondary receives almost pure helium matter of 
$\Delta M_{\rm He} \sim 0.1-0.5 M_\odot$ 
to form a helium-enriched envelope.
The secondary's hydrogen content decreases to 
$X \sim 0.6$ if helium is completely mixed into the star.  
For the $9+2.5 M_\odot$ case, much more helium is transferred,
but much less helium for the $6+2 M_\odot$ case.

\section{Shrink of orbital separation during superwind phase}
     Based on the population synthesis analysis, 
Yungelson \& Livio (1998) claimed that almost no realization frequency
is derived for the original WD+RG model (Hachisu et al. 1996). 
In this section, we point out that very wide binaries with initial 
separations of $a_i >~ 1500 ~R_\odot$, which were not included in their
analysis, are essentially important in our SN Ia modeling.
\par
     The more massive component (the initial mass of $M_{1,i}$) of 
a binary first evolves to a red giant (AGB stage) and fills its inner 
critical Roche lobe.  After a common envelope phase, 
the more massive component leaves a C+O WD, 
and the separation of the binary decreases by a factor of 
$a_f/a_i  \sim 1/40 - 1/50$ from eq.(2)
for $M_{\rm WD} \sim 1 ~M_\odot$ 
and $M_2 \sim 1 ~M_\odot$ because a $\sim 1 ~M_\odot$ WD descends 
from a main-sequence star of $M_{1,i} \sim 7-8 ~M_\odot$.
Yungelson \& Livio (1998) assumed that 
the separation of interacting binaries is {\it smaller than} 
$a_i <~ 1500 ~R_\odot$. 
Then, the widest binaries have separations of 
$a_f <~ 30-40 ~R_\odot$  after common envelope evolution.  
The orbital period is 
$P_0 <~ 20$ days for $M_{\rm WD,0} \sim 1 ~M_\odot$ 
and $M_{\rm RG,0} \sim 1 ~M_\odot$.  
There is no SN Ia region of the WD+RG systems 
for $P_0 <~ 20$ days as seen from Fig. 6. 
Thus, Yungelson \& Livio concluded that we cannot expect 
any SN Ia explosions from the right-hand SN Ia region 
(WD+RG system) in Fig. 6.
\par
     The reason that Yungelson \& Livio's modeling fails to reproduce
our SN Ia progenitors is their assumption of $a_i <~ 1500 ~R_\odot$. 
In what follows, we show that WD+RG binaries wide enough 
to have $P_0 \sim 100-1000$ d are born from initially very wide 
binaries with $a_i \sim 1500-40000 ~R_\odot$.
\par
     A star with a zero-age main-sequence mass of 
$M_{1,i} < 8 ~M_\odot$ ends its life by ejecting its envelope in a
wind of relatively slow velocities ($v \sim 10-40$ km s$^{-1}$) 
as shown in Fig. 5.  
These wind velocities are as low as the orbital velocities of 
binaries with separations of $a_i \sim 1500-40000 ~R_\odot$
for a pair of $M_{1,i} \sim 7 ~M_\odot$ and $M_{2,i} \sim 1 ~M_\odot$.
When the wind velocity is as low as the orbital velocity, 
we have the separation shrinking due to effective loss of
the angular momentum by the outflowing gas, i.e., 
$\dot a/a \approx 2 \dot M_1 / M_2 < 0$ (see Hachisu et al. 1999a).
\par
     Once the binary system begins to shrink, its evolution becomes 
similar to a common envelope evolution 
(see {\it stages B-D} in Fig. 5). 
Thus, the separation is reduced by a factor of $1/40-1/50$, i.e., 
$a_f \sim 30-1000 ~R_\odot$ 
for $M_{1,i} \sim 7 ~M_\odot$ and $M_{2,i} \sim 1 ~M_\odot$.
The orbital period becomes $P_0 \sim 15-3000$ days for a pair of
$M_{\rm WD,0} \sim 1 ~M_\odot$ and $M_2 \sim 1 ~M_\odot$
({\it stage D} in Fig. 5).
These initial sets of the parameters are very consistent with 
the initial conditions of our WD+RG progenitor systems.

\begin{figure}
\plottwo{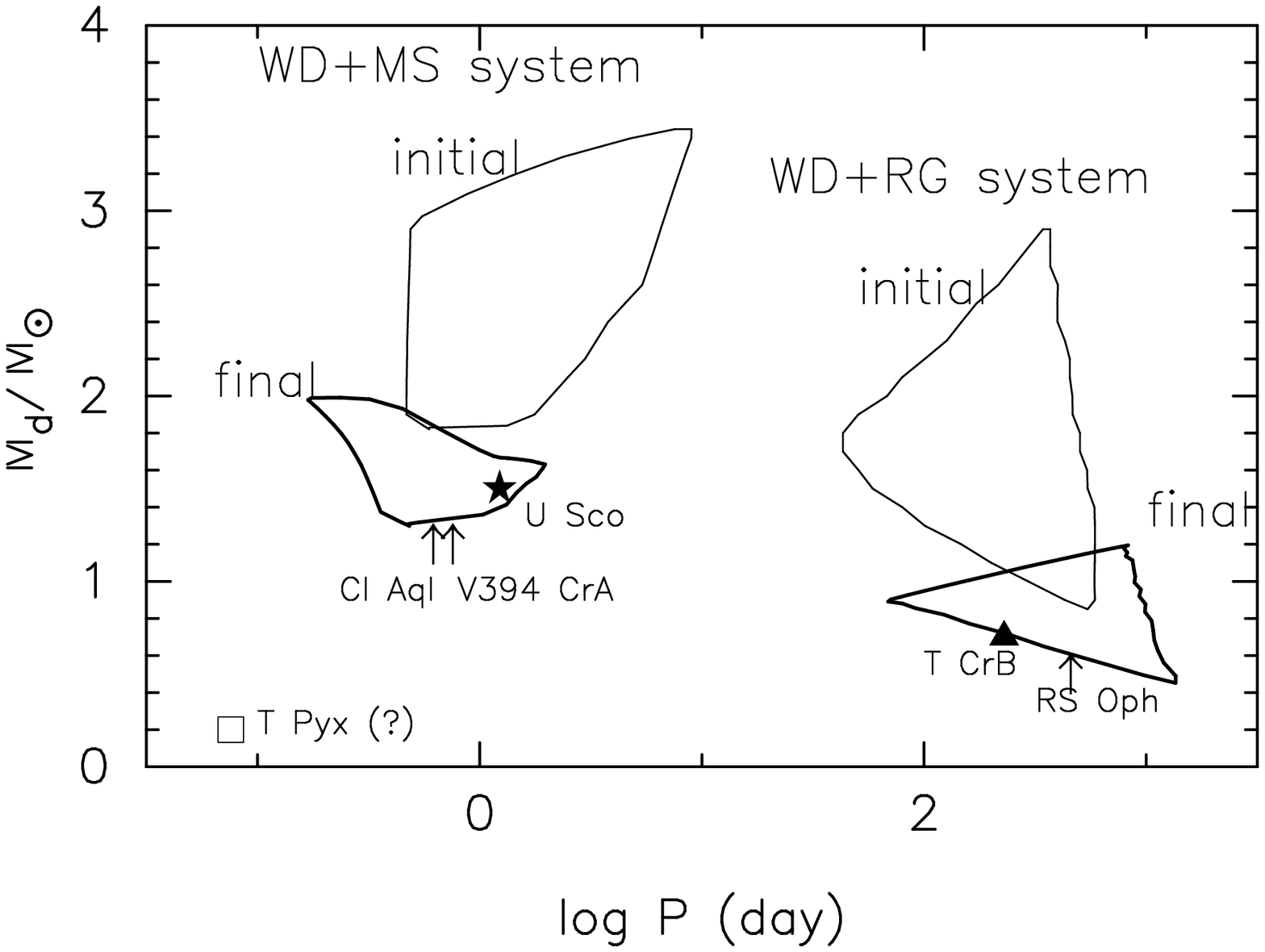}{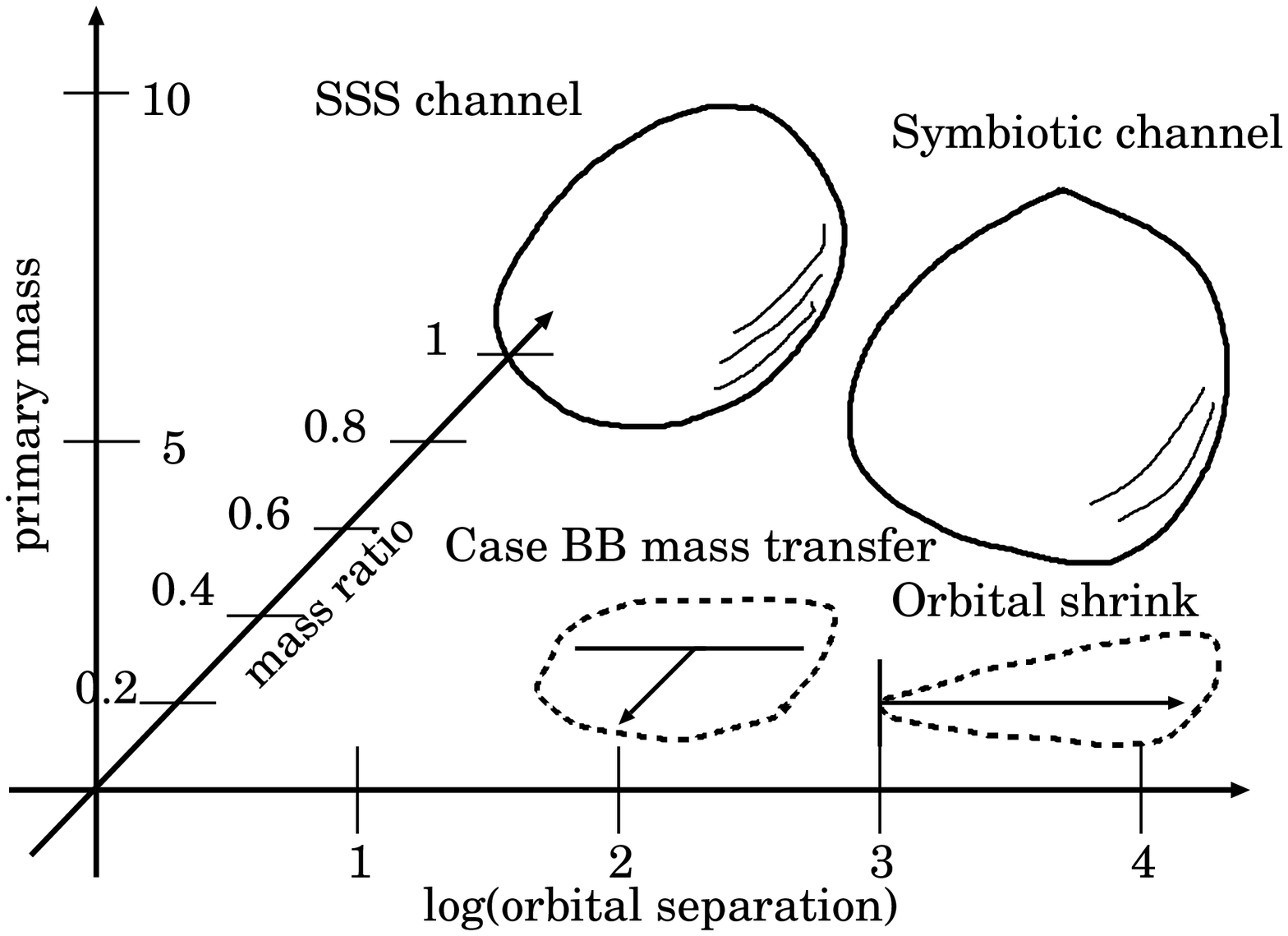}
\caption{
{\bf Left:} 
Regions producing SNe Ia are plotted  
in the $\log P - M_{\rm d}$ (orbital period --- donor mass) plane 
for the WD+MS system ({\it left}) and the WD+RG system ({\it right}).
Here we assume an initial white dwarf mass of $1.0 M_\odot$. 
Currently known positions of each recurrent nova are indicated 
by a star mark ($\star$) for U Sco (Hachisu et al. 2000),
a filled triangle for T CrB (Belczy\'nski \& Mikolajewska 1998),
an open rectangle for T Pyx (Patterson et al. 1998),
but by arrows for the other three recurrent novae, V394 CrA,
CI Aql, and RS Oph,
because the mass of the companion is not yet available explicitly.
{\bf Right:}
The accretion winds of WDs open new two channels to SNe Ia
while the two evolutionary processes of the case BB mass transfer
and the orbital shrinkage during the superwind phase enlarge 
the initial phase volume of SN Ia progenitors 
for the SSS channel and the symbiotic channel, respectively. 
}
\end{figure}

\section{Type Ia supernova rates}
     We estimate the SN Ia rate from the WD+RG and WD+MS systems 
in our Galaxy by 
\begin{equation}
\nu = 0.2 \int d q \int d~\log A 
\int {{d M} \over M^{2.5}} 
\quad {\rm yr}^{-1},
\end{equation}
where $q$, $A$, and $M$ are the mass ratio ($q=M_2/M_1 \le 1$),
the initial separation in units of solar radius, 
and the primary initial mass
in solar mass units (from eq. [1] of Iben \& Tutukov 1984).
The SN Ia rate is essentially the same as the volume of
the initial phase space consisting of $q$, $A$, and $M$ 
as shown in Fig. 6.  
\par
     The accretion winds of WDs open new two channels to SNe Ia.
The case BB mass transfer in the SSS channel extends 
the phase volume towards the low mass ratio region as shown
in Fig. 6.  On the other hand, the orbital shrinkage 
during the superwind phase in the symbiotic channel enlarges
the phase space towards the large initial separation
also shown in Fig. 6.
Therefore, the estimated rate of WD+RG/WD+MS systems becomes close 
to the observed rate in our Galaxy, i.e., $\nu \sim 0.003$ yr$^{-1}$,
which comes from the SSS channel of $\nu_{\rm MS} \sim 0.001$ yr$^{-1}$
and the symbiotic channel of $\nu_{\rm RG} \sim 0.002$ yr$^{-1}$.

\end{document}